# Superconductivity up to 35 K in the iron-platinum arsenides (CaFe$_{1-x}$Pt$_x$As)$_{10}$Pt$_{4-y}$As$_8$ with layered structures

*Catrin Löhnert, Tobias Stürzer, Marcus Tegel, Rainer Frankovsky, Gina Friederichs, Dirk Johrendt\**

The discovery of high-$T_c$ superconductivity in iron arsenides in 2008[1] has arguably been the biggest breakthrough in this field since the appearance of the copper oxide superconductors in 1986. In iron arsenides, superconductivity up to 55 K[2] originates in layers of edge-sharing $^2_\infty$[FeAs$_{4/4}$]-tetrahedra. Meanwhile, a series of different structure types have been identified, but the family of superconducting iron arsenide compounds is still small in comparison with the cuprates. Its members are mainly derivatives of the relatively simple and long known *anti*-PbFCl-[1,3,4] and ThCr$_2$Si$_2$-type structures.[5,6] Thus extending the crystal chemistry of iron based superconductors is a foremost task of solid state chemistry. Compounds like Sr$_2$VO$_3$FeAs with thick perowskite-like oxide-blocks between the FeAs-layers were derived from known copper sulphides and showed superconductivity up to 37 K.[7] However, the combination of the metallic iron arsenide layers with transition metal oxides caused difficulties.[8, 9]

Another approach is the combination of iron arsenide layers with other intermetallic building blocks. We considered the fact that a second transition metal should be one that can adopt coordination geometries other than tetrahedral. Platinum appeared promising, because Pt is known to be very flexible and forms arsenides with octahedral, tetrahedral, trigonal and square coordination in compounds like PtAs$_2$,[10] SrPt$_2$As$_2$,[11] SrPtAs,[12] and Cs$_2$PtAs$_2$,[13] respectively. Recently, Nohara *et al.*[14] mentioned superconductivity in the system Ca-Fe-Pt-As, but the structure and composition of the compound have not been reported. With these points in mind, we started explorative syntheses in the system Ca-Fe-Pt-As and found three new platinum-iron arsenides (CaFe$_{1-x}$Pt$_x$As)$_{10}$Pt$_4$As$_8$ (CaFe$_{1-x}$Pt$_x$As)$_{10}$Pt$_3$As$_8$ and (CaFeAs)$_{10}$Pt$_4$As$_8$. These compounds crystallize in so far unknown structure types, where iron arsenide and platinum arsenide layers alternate. We have detected superconductivity up to 35 K, which is probably either induced by Pt-doping of the FeAs-layers in (CaFe$_{1-x}$Pt$_x$)$_{10}$Pt$_3$As$_8$ or by indirect electron doping in (CaFeAs)$_{10}$Pt$_4$As$_8$ due to additional Pt$^{2+}$ in the platinum arsenide layers. However, the concrete phase relationships are not yet completely resolved. In this communication, we report the synthesis, crystal structures, preliminary property measurements and DFT calculations of these new superconductors.

[*] Catrin Löhnert, Tobias Stürzer, Rainer Frankovsky, Gina Friederichs, Dr. Marcus Tegel, Prof. Dr. Dirk Johrendt,
Department Chemie
Ludwig-Maximilians-Universität München
Butenandtstr. 5-13 (Haus D), 81377 München
Fax: +49 (0)89 2180 77431
E-mail: johrendt@lmu.de

[**] This work was financially supported by the German Research Foundation (DFG) within the priority program SPP1458, project JO257/6-1

Supporting information for this article is available on the WWW under http://www.angewandte.org or from the author

The polycrystalline samples were mostly inhomogeneous and contained plate-shaped and well as needle-shaped crystals with metallic lustre. X-ray powder patterns could initially not be indexed, and the plate-like crystals easily exfoliated. Their diffraction patterns showed clean square motifs, but a disturbed periodicity of the third dimension which indicated stacking disorder. Only some crystals were of fairly sufficient quality for x-ray structure determinations. Finally, we found three different, but closely related crystal structures, whose compositions and lattice parameters are compiled in Table 1. The rather high $R$-values reflect the poor crystal quality and still not completely resolved twinning and/or intergrowth issues. For detailed crystallographic data we refer to the supplementary material.

**Table 1.** Crystal data of the platinum-iron arsenides

| formula | (CaFe$_{1-x}$Pt$_x$As)$_{10}$Pt$_3$As$_8$ | (CaFe$_{1-x}$Pt$_x$As)$_{10}$Pt$_{4-y}$As$_8$ | |
|---|---|---|---|
| label | 1038 | α-1048 | β-1048 |
| SG | $P\bar{1}$ | $P4/n$ | $P\bar{1}$ |
| $a$ /Å | 8.776(1) | 8.716(1) | 8.7382(4) |
| $b$ /Å | 8.781(1) | $a$ | 8.7387(4) |
| $c$ /Å | 10.689(2) | 10.462(2) | 11.225(1) |
| $\alpha$ /° | 75.67(1) | 90 | 81.049(3) |
| $\beta$ /° | 85.32(1) | 90 | 71.915(3) |
| $\gamma$ /° | 89.97(1) | 90 | 89.980(3) |
| $R_{Fo>3\sigma Fo}$ | 0.064 | 0.099 | 0.069 |

(CaFe$_{1-x}$Pt$_x$As)$_{10}$Pt$_3$As$_8$, hereinafter referred to as the 1038-compound, crystallizes triclinic and consists of iron arsenide layers separated by calcium ions and slightly puckered planar Pt$_3$As$_8$-layers with platinum in square coordination of arsenic. The crystal structure is shown in Figure 1.

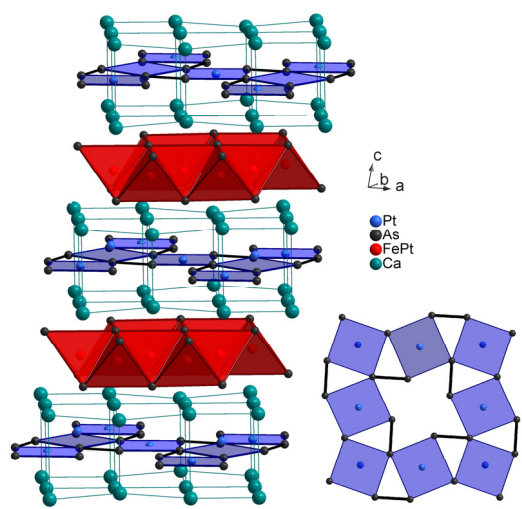

**Figure 1.** Crystal structure of triclinic (CaFe$_{1-x}$Pt$_x$As)$_{10}$Pt$_3$As$_8$ (1038-compound), and details of the Pt$_3$As$_8$-layer (right).

Platinum atoms are either lie in plane in or slightly shifted above the corner-sharing $As_4$-squares. Arsenic forms $As_2^{4-}$ dumbbells with typical As-As bond lengths of ≈ 2.48 Å. The combination of $As_2$-dumbbells with square coordinated Pt-atoms is known from the pyrite-like compound $BaPt_4As_6$, which also contains octahedrally coordinated platinum.[15] Assuming divalent $Pt^{2+}$ ($5d^8$) in the present compound, charge neutrality is achieved according to $(Ca^{2+}Fe^{2+}As^{3-})_{10}Pt^{2+}{}_3[(As_2)^{4-}]_4$. Thus the electronic situation of the $(FeAs)^{1-}$ layer is identical to the known parent compounds $BaFe_2As_2$ and LaFeAsO. Subsequent refinements of the crystal structure revealed Pt-substitution at the Fe-site. The final composition has been determined to be $(CaFe_{0.95(1)}Pt_{0.05(1)}As)_{10}Pt_3As_8$.

A second type of plate-like crystals from the polycrystalline samples showed tetragonal symmetry and the structure could be solved in the space group $P4/n$. The tetragonal structure contains building blocks very similar to the triclinic phase as described above. No Pt-doping was detected at the iron site, even though the refined composition $(CaFeAs)_{10}Pt_{3.58(2)}As_8$ (α-1048) contains even more platinum than triclinic $(CaFe_{0.95(1)}Pt_{0.05(1)})_{10}Pt_3As_8$. Finally also the structure of the needle-shaped crystals could be solved and refined in the space group $P\bar{1}$. Their composition is $(CaFe_{0.87(1)}Pt_{0.13(1)}As)_{10}Pt_4As_8$ (β-1048), which is the platinum-richest phase so far. In this case, all Pt-sites in the $Pt_4As_8$-layer are fully occupied, and additionally about 13% iron in the FeAs- layer is substituted by platinum.

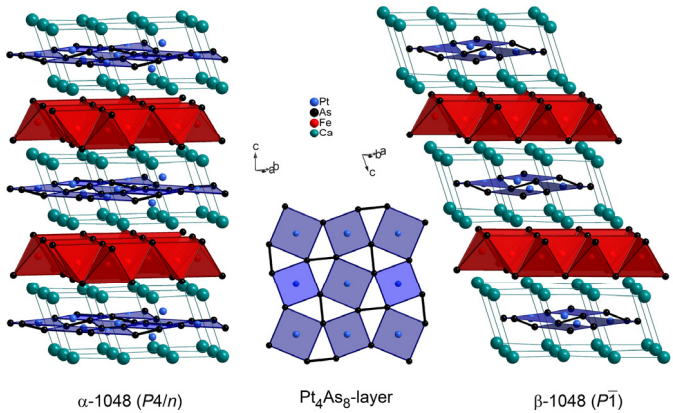

**Figure 2.** Crystal structures of $(CaFeAs)_{10}Pt_{3.7}As_8$ (left, α-1048) and $(CaFe_{0.87(1)}Pt_{0.13(1)}As)_{10}Pt_4As_8$ (right, β-1048). The $Pt_4As_8$-layer of both structures is shown in the middle.

Fig. 2 shows the crystal structures of both 1048-compounds. Additional Pt-atoms occupy the voids in the $Pt_3As_8$-layer of the 1038-structure (compare Fig. 1), which gives a layer formula of $Pt_4As_8$. But also the stacking of $Pt_4As_8$- and FeAs-layers is different from the 1038-compound. This becomes clear by comparing the arrangements of the calcium atoms in the 1038- and 1048-structures. While the Ca-layers are mirror-symmetric above and below the $Pt_3As_8$-layers in the 1038-compound (see Fig. 1), they are shifted by half the diagonal of one CaFeAs-subcell in the 1048-structures. On the other hand, the FeAs-layers are congruently stacked in the 1048-structures, but not in the 1038-structure. Thus, α- and β-1048 are polymorphs with the same stacking of the Ca- and FeAs-layers. But while consecutive $Pt_4As_8$-layers are congruent in α-1048, they are shifted by one period of the CaFeAs-layer (3.89 Å) along [120] in β-1048 (Fig. 2). This shift by (0.4,0.2,0) is incompatible with the positions of the 4-fold axis at $(0,0,z)$ and $(½,½,z)$, therefore the structure becomes triclinic.

The crystal structures are in agreement with the x-ray powder patterns that could be fitted by using both the 1038- and the α-1048-phases in roughly 60:40 weight ratio. Figure 3 shows the measured data and the Rietveld fit.

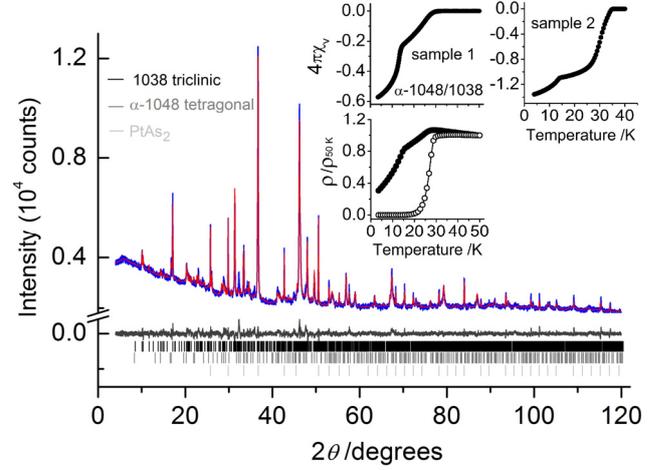

**Figure 3.** X-ray powder pattern and Rietveld-fit using the structures of the triclinic 1038- and the tetragonal α-1048-compounds. Inset: AC-susceptibility and DC-resistivity measurements. The latter shows data of a cold-pressed pellet (filled circles), and after annealing at 1000°C (open circles).

AC-susceptibility measurements of different samples always revealed two superconducting transitions. A lower onset temperature around 15 K was found in both samples, while the higher $T_c$ is at 31 K in sample 1 and at 35 K in sample 2, respectively (insets in Fig. 3). The shielding fraction of about 60% at 3 K in sample 1 can be subdivided in two roughly equal amounts of the two phases, which is in agreement with the Rietveld data, but allows no assignment of the different phases to the superconducting transitions. The resistivity of a cold pressed pellet of sample 1 shows the same two transitions as the susceptibility, but zero resistivity is not achieved due to grain boundary effects. After annealing the pellet at 1000°C for five hours, we observe one rather sharp drop at ~30 K (open circles in Fig. 3), while the magnetic susceptibility of the annealed pellet (not shown) still reveals two transition temperatures. As expected, the fraction of the 30 K superconductor is sufficient to achieve zero resistivity.

Both compounds show electronic structure features typical for iron arsenide superconductors that may allow an at least probable assignment of the 1038- and 1048-compounds to the observed transitions. The 1038-phase formally represents a parent compound with $FeAs^{1-}$-layers like undoped $BaFe_2As_2$ or LaOFeAs, both non-superconducting. Pt-doping at the Fe-site is known to induce superconductivity in $SrFe_{2-x}Pt_xAs_2$,[16] and we suggest that $(CaFe_{1-x}Pt_xAs)_{10}Pt_3As_8$ is superconducting because of Pt-doping of the FeAs-layers. On the other hand, indirect electron-doping of clean FeAs-layers induces superconductivity in $LaO_{1-x}F_xFeAs$ and $Ca_{1-x}RE_xFe_2As_2$[17]. The tetragonal α-1048-compound may be considered as indirectly electron-doped due to ≈0.6 additional $Pt^{2+}$-atoms in the $Pt_{3.58}As_8$-layer, which formally reduces the charge at the Fe atom. It has generally been observed that indirect electron- or hole-doping of clean FeAs-layers leads to higher $T_c$'s than substitution of Fe by other metals (direct doping), for example $Ba_{0.6}K_{0.4}Fe_2As_2$ (38 K)[6, 18] and $BaFe_{1.86}Co_{0.14}As_2$ (22 K)[19]. One possible reason may be the additional disorder in the latter case.

From these considerations, we suggest that the indirectly electron-doped α-1048-compound with clean FeAs-layers has the higher $T_c$ than the Pt-doped 1038-compound. We have also synthesized samples of the 1038-phase with lower Pt-concentrations (no Pt at the Fe-site) that were non-superconducting, which is in line with our arguments.

Our assignment is compatible with the properties of the β-1048-compound. Due to the needle-like shape of the crystals, we were able to manually select an amount sufficient for AC-measurements and powder diffraction, but not for resistivity measurements. The pattern was fitted using the crystal structure of β-1048 (Fig. 4). Small amounts of $PtAs_2$ were included in the refinement. The peak at 2θ = 16° is a still unknown impurity, but nevertheless the pattern is well described by the structure of the β-1048 obtained from x-ray single crystal data.

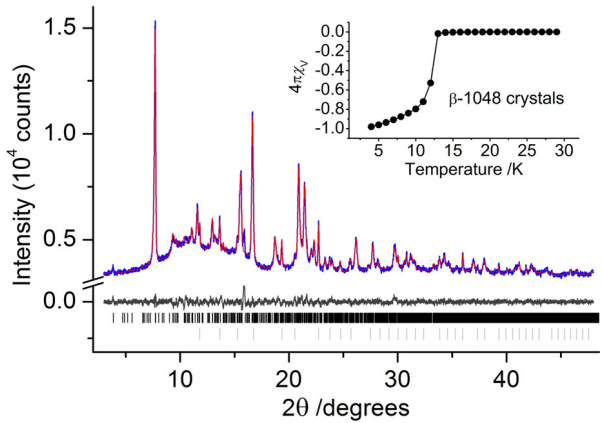

*Figure 4.* X-ray powder pattern and Rietveld-fit of the needle-shaped β-1048 crystals. Inset: AC-susceptibility measurement.

The AC-susceptibility of the crystals is shown in the inset of Fig. 4. A sharp superconducting transition at 13 K with almost 100% shielding at 4 K is observed in agreement with the single phase refinement of the x-ray powder data. Due to the tiny amount of these crystals, this 13 K-transition is not visible in the AC-measurement and in the powder pattern of the whole sample (Fig. 3). The lower critical temperature is plausible, because the crystals of the β-1048 compound can be considered as overdoped. Indeed, Nishikubo *et al.*[16] observed superconductivity at 17 K in $Sr(Fe_{1-x}Pt_x)_2As_2$ at 12.5% Pt-doping. Our β-1048 single crystals contain about the same amount of Pt at the iron-site (13%) and are additionally indirectly electron-doped due to the completely Pt-filled $Pt_4As_8$-layer.

DFT calculations were conducted in order to check for certain features of the electronic structure that were considered essential. In FeAs superconductors, the electronic states near the Fermi-level ($\varepsilon_F$) are dominated by iron 3$d$-bands. These generate a special topology of the Fermi surface, referred to as nesting between so-called electron- and hole-like sheets.[20] It has been argued that this nesting plays a key role in the pairing mechanism,[21] however, the recently discovered iron selenide superconductors gave rise to doubts about this concept.[22]

The question that arises here is whether the electronic subsystem of the $Pt_{4-y}As_8$-layer contributes to the Fermi surface. If not, we can probably apply the concept of the other iron arsenide materials; otherwise a different scenario has to be considered. From the chemical point of view, one may expect that the Pt-5$d_{x^2-y^2}$ orbitals are pushed above the Fermi-level by the square ligand field. If the Fermi level in the FeAs bands is just inside this gap, we have a pure FeAs Fermi surface. The 1038-compound is charge-neutral by using the Zintl concept according to $(Ca^{2+}Fe^{2+}As^{3-})_{10}Pt^{2+}_3[(As_2)^{4-}]_4$, while two additional electrons have to be placed in the 1048 structure that may be written as $(Ca^{2+}Fe^{2+}As^{3-})_{10}Pt^{2+}_4[(As_2)^{4-}]_4 \cdot 2e^-$. However, it is not yet clear where to ascribe these extra electrons. Fig. 5 shows the partial density of states (PDOS) of the Pt-5$d$ and Fe-3$d$ orbitals. We also show the Crystal Orbital Hamilton Populations (COHP) of the Pt-As bonds in the $Pt_{4-y}As_8$-layers that provide information about the bonding/antibonding characters of the respective electronic states.

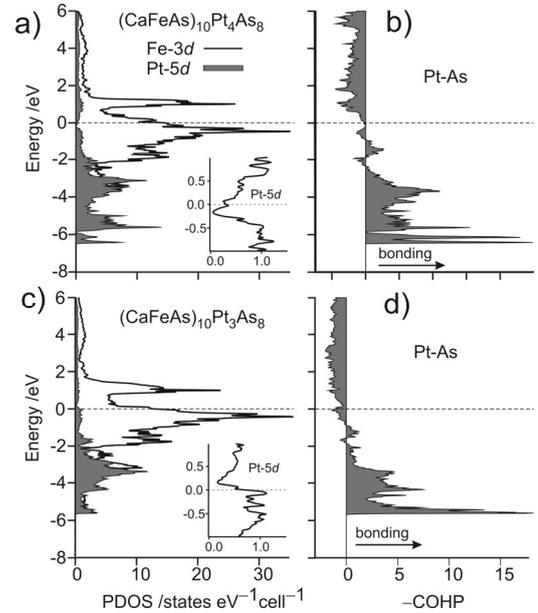

*Figure 5.* Partial Density of States (PDOS) and crystal orbital Hamilton population (COHP) of the Pt-As bonds in tetragonal α-1048 (a,b) and triclinic 1038 (c,d). Insets: Details of the Pt-5$d$ PDOS near $\varepsilon_F$.

The Fe-3$d$ PDOS (Fig. 5 a+c) of both compounds are very similar to those of other FeAs superconductors. The Fermi-levels at the rising edges of the Fe-3$d$ peaks reveal that mostly iron states contribute to the Fermi-surfaces in both the α-1048 and 1038-compounds. On the other hand, the contribution of platinum at $\varepsilon_F$ is very small. The Pt-5$d$ PDOS of α-1048 has nearly a gap, which means that the Fermi surface of this compound consists of states from the FeAs-layer only, as in the known FeAs superconductors. This is also consistent with the Pt-As bonding. The COHP plot (Fig. 5b) reveals that the majority of the Pt-As bonding states are well below and the antibonding states are mainly well above the Fermi-energy, respectively. In other words, Pt-As bonding removes most of the Pt-states from the Fermi-surface in α-1048. The situation is surprisingly similar in the 1038-compound (Fig. 5 c+d). The Pt-contribution at $\varepsilon_F$ again is very small, while these states are slightly Pt-As antibonding (Fig 5d). Details of the Pt-5$d$ PDOS are shown in the insets of Fig. 5. The Fermi level is just above the gap in the α-1048-compound (which contains one more $Pt^{2+}$), but just below this gap in the 1038-phase. Thus band filling across this gap in the Pt-states mainly fills Fe-states that contribute most of the energy levels in this range, which is equivalent to electron doping. This strongly suggests that the FeAs-layer of the α-1048-compound is indirectly doped by 2 electrons from the $Pt_4As_8$-layer. Assuming $Pt^{2+}$, the amount of transferred charge is $0.2e^-$/FeAs, which is close to the typical values where other indirectly electron-doped iron-arsenide superconductors like $LaO_{1-x}F_xFeAs$[1] or the recently discovered $Ca_{1-x}RE_xFe_2As_2$[17] achieve the highest critical temperatures.

In summary, we have found three new superconducting iron-platinum arsenides $(CaFe_{1-x}Pt_xAs)_{10}Pt_{4-y}As_8$. The crystal structures are stacking variants of FeAs- and slightly puckered $Pt_{4-y}As_8$-layers with square coordinated platinum, separated by calcium-layers, respectively. Arsenic atoms in the $Pt_{4-y}As_8$-layers form $As_2^{4-}$ dumbbells according to Zintl's concept, providing charge balance in $(Ca^{2+}Fe^{2+}As^{3-})_{10}Pt^{2+}_3[(As_2)^{4-}]_4$.

Superconductivity was observed at 13-35 K. We suggest that the highest $T_c$ above 30 K occurs in the α-1048 phase with clean FeAs-layers that are indirectly electron-doped according to $(Ca^{2+}Fe^{2+}As^{3-})_{10}Pt^{2+}_4[(As_2)^{4-}]_4 \cdot 2e^-$. We also suggest that the lower critical temperatures appear in the 1038- and β-1048 phases due to Pt-doping at the Fe-site. Such direct electron-doping has not achieved $T_c$ above 25 K in any other iron-based material.

DFT band structure calculations suggest that the contribution of the $Pt_{4-y}As_8$-layers to the Fermi surface is small and the Fermi energy is slightly either below or above a quasi-gap in the Pt-states of the 1038- and α-1048 compounds, respectively. The latter clearly supports the suggested indirect electron-doping of the FeAs-layer in the α-1048 compound with the highest critical temperature. The platinum-iron compounds represent the first iron-based superconductors with new crystal structures and can serve as a new platform for further studies that go beyond the known systems.

*Note added:*
During the submission of this manuscript, we noticed a preprint by Ni *et al.*[23] that reports similar results. The authors confirm two of the crystal structures reported here and observed superconductivity up to 27 K.

## Experimental Section

The new compounds were synthesized by heating of stoichiometric mixtures of pure elements at 700-1100 °C in alumina crucibles, sealed under argon in silica tubes. Powder diffraction data were measured using either a Huber G670 Guinier imaging plate (Co-K$_{α1}$ or Cu-K$_{α1}$-radiation) or a STOE Stadi P diffractometer (Mo-K$_{α1}$ or Cu-K K$_{α1}$-radiation, Ge-111 monochromator). Rietveld refinements were performed with the TOPAS package.[24] Crystals were selected from the polycrystalline samples. X-ray intensity data were measured with a STOE IPDS-I imaging plate or a Nonius-κ-CCD (Mo-Kα-radiation). The structures were solved using the charge flipping method included in the Jana2006 program package.[25] The latter was also used for structure refinement. Details of the crystal structure determinations in CIF-format files are available as supplementary material. AC-susceptibility data were measured at 1333 Hz and 2G. DFT calculations were performed with the LMTO47c package.[26]